# Willem Janszoon Blaeu

Jean-Pierre Luminet
Laboratoire Univers et Théories, CNRS-UMR 8102,
Observatoire de Paris, F-92195 Meudon cedex, France
E-mail: `jean-pierre.luminet@obspm.fr`

This article describes the life and work of Willem Janszoon Blaeu (1571-1638), who founded one of history's greatest cartographic publishing firms in 1599. Mostly renowned as a cartographer, he also made terrestrial and celestial globes, various instruments such as quadrants, a planetarium and a tellurium. He invented mechanical devices for improving the technics of printing. As an astronomer, a former student of Tycho Brahe, Willem Blaeu made careful observations of a moon eclipse, he discovered a variable star now known as P Cygni, and carried out a measurement of a degree on the surface of the earth (as his countryman Snell did in 1617).

The Blaeu family has its origin in the island of Wieringen, where about 1490, Willem Jacobszoon Blauwe - the grandfather of Willem – was born. From his marriage with Anna Jansdochter sprang six children. The second son, Jan Willemsz. (1527- before 1589) was the father of Willem Blaeu, and continued the family tradition by practising the prosperous trade of herring packer. From his second marriage with Stijntge, Willem Jansz. Blaeu was born at Alkmaar or Uitgeest.

At an early age, Willem Blaeu went to Amsterdam in order to learn the herring trade, in which he was destined to succeed his father. But Willem did not like this work very much, being more inclined to Mathematics and Astronomy. He did not attend a university and worked first as a carpenter and a clerk in the Amsterdam mercantile office of his cousin Hooft.

However, in 1595 he became a student of Tycho Brahe (1546-1601). The celebrated Danish astronomer demanded a high standard of his pupils. Some were invited by him, others were undoubtedly taken on special recommendation. We may therefore presume that young Blaeu had reached a good standard of education and technical skill, since he was considered worthy to become a student of the great astronomer. Blaeu lived on the Island of Hven over the winter of 1595/1596, at Brahe's famous observatory in Uraniborg. Thanks to this exact knowledge acquired from Brahe, Blaeu was able to make tables for sun declination ; especially he also learned from Brahe to make globes and instruments like the quadrants.

As it is well-known, Tycho Brahe had his own cosmic system, a sort of compromise between the Ptolemaic and Copernican. Willem Blaeu, although a supporter of the Copernican system, remained cautious during the rest of his career. In his books he mentioned the Copernican model as one of the existing theories, besides the Ptolemaic and Tychonic. It will not only save him for confrontations with religious people, but this attitude was also beneficial for his sales.

After his return from Hven in 1596, Blaeu settled in Alkmaar. Very little is known of his stay here. He married, probably in 1597, Marretie or Maertgen, daughter of Cornelis from Uitgeest. Here too, his eldest son Joan was born.

In Alkmaar, on 21 February 1598, Blaeu observed an eclipse of the moon, which was also seen by Tycho Brahe at Wandsbeck, near Hamburg. They made their observations in accordance in order to determinate the difference of longitude between the two places. The following year Blaeu made for Adriaan Anthonisz. a 34 cm. diameter globe, based on Brahe's as yet unpublished information.

Blaeu moved in 1598/9 from Alkmaar to Amsterdam, where he soon established himself as a merchant of maps and globes, an instrument maker and a printer, while continuing to perform some astronomical observations.

As a producer of globes, following the tradition Blaueu made his globes in pairs : a terrestrial and a celestial one. After his first globe of 1599, Blaeu produced in 1602 a small 23.5 cm globe which he dedicated to the States of Holland, Zeeland and West-Friesland. In 1603, he introduced the southern constellations on a celestial globe (his great rival Jocodus Hondius had been the first to do this two years earlier). Blaeu's 68 cm globes were made in response to the 53.5 cm pair issued by the Hondius firm in 1613. They were presented in 1616 to the States General, who awarded an honorarium of 50 guilders. They would remain the largest globes in production for over 70 years, until Vincenzo Maria Coronelli (1650-1717) issued his 110 cm pair in 1688. In 1634 he published an important manual for making globes and sundials, *Tweevoudigh onderwijs van de Hemelsche en Aerdsche globen* (Twofold instruction in the use of the celestial and terrestrial globes), often reedited.

As an astronomer, in 1600 Blaeu was the first to note a third magnitude star in a place where no star had been recorded before, and he made well-documented observations of it. On a globe made also by Blaeu (now in a Prague museum), it is written: "The new star in Cygnus that I first observed on August 8, 1600, was initially of third magnitude. I determined its position by measuring its distance from Vega and Albireo. It remains in this position but now is no brighter than 5th magnitude." It was indeed the third variable star to be discovered : over the next few years, the star faded below naked-eye visibility, but returned to magnitude 3.5 in 1655, where it remained until 1659. Today we know that P Cygni (also known as 34 Cygni) is a Be variable star, one of the most luminous stars known, which lies about 7,000 light-years away.

In 1617, Willebrord Snellius (1580–1626), a countryman and collaborator of Willem Blaeu, published in Leyden his work *Eratosthenes Batavus* (The Dutch Eratosthenes), where he described the method and gave the result of his geodetic operations between Alkmaar and Bergen op Zoom – two towns separated by one degree of the meridian, which he measured to be equal to 117,449 yards (107.395 km). For this he used a huge quadrant (with a radius of over 2 meters) of wood with a brass mounting made by Willem Blaeu after the example of Tycho Brahe's large quadrant. Soon after a similar operation was undertaken in the same region by Blaeu himself ; it appears to have been executed with a great accuracy, but the details have never been published.

As an instrument maker, Blaeu had enjoyed an excellent training under Tycho Brahe. In the sixteenth century, the art of making instruments flourished especially in the Southern Netherlands. As a result of the great discoveries navigation advanced, and there was a need for astronomical instruments to determinate positions. In his sea atlases, Blaeu showed considerable interest in the instruments used at sea, and illustrated or reproduced them by means of movable diagrams.

Blaeu devoted his attention to the needs of navigation from an early stage of his career. His first publication in this field was his *Nieuw graetbouck*. Blaeu issued two pilot guides for the description of the Eastern, Western and Northern navigation, named *Het licht der Zee-vaert*

(first edition 1609 ; English translation : *The light of navigation*, 1620) and *Zeespiegel* (first edition 1623). The works were republished several times : in the history of early Dutch pilot guides, Blaeu's work takes a very important place.

As a printer, Willem Jansz. Blaeu made substantial improvements in the moving parts of the printing press. About 1620 in Amsterdam, he added a counterweight to the pressure bar in order to make the platen rise automatically ; this was the so-called « Dutch press », the design of which became almost general throughout the low contries and were introduced to England ; a copy was to be the first press introduced into North America in 1639.

Blaeu employed the best pressman, engravers, scibes and colorists. His types were clean and well cut; his paper, bearing his own watermark, was heavy and of good quality. He published works including that of famous Dutch writers like P.C. Hooft and Joost van den Vondel. His printing had an international high reputation. Blaeu was prepared to publish everyone's work. Although he had Remonstrant sympathies, he printed books for Catholics, Jews and various Protestant groups. Counter-remonstrants, Remonstrants, Baptists, Socinians or dissenters, as for printing it made no difference for Willem. However, because of caution he published under another name and Cologne was given as place of issue. Because of his large publisher's list he opened an extra printery on the Bloemgracht in 1635. There the Blaeu establishment boasted nine flat-bed presses for letter press printing, six presses for copperplate printing and a typefoundry in 1664. His tolerance provided him at least a large income.

But it is as a cartographer that Willem Blaeu was mostly celebrated. For instance, his wall maps are considered to be among the most influential and artistically virtuous masterpieces of the great era of baroque cartography. The publication of the first set of his wall maps in 1608 was responsible for initiating his ascendancy to the preeminent position in the highly competitive global map market. Blaeu published several wall maps printed on parchment or paper. The use of maps as wall hangings in contemporary Dutch houses went beyond the desire for cartographic information. Maps were also used to express status, to promote a better understanding of history or politics or to take the place of paintings. At the moment Vermeer painted his works, Amsterdam was the world centre of map-making. Among the most majestic productions were the wall maps of Willem Blaeu. The pride of place that wall maps claimed in Dutch homes is most eloquently presented in the exquisite paintings of Vermeer. These views of everyday life bear witness to an almost totemic cult of maps.

In 1605, Willem Blaeu moved to the nowadays called Damrak, where most of the Amsterdam booksellers and mapmakers were established. On Damrak, at that time a canal in the centre of Amsterdam, they had direct contact with sailors. The Blaeu's house was called "In de Vergulde Sonnewijser" (In the gilt sundail). By 1608, Willem Blaeu had already published a fine world map and a popular marine atlas. He then began planning a major atlas intended to include the most up-to-date maps of the entire world. Progress was extremely slow, and although he spent the rest of his life compiling maps for this ambitious project, the atlas was completed well after his death by his son Joan – see below.

Next to Blaeu's shop were Johannes Janssonius and Hondius' houses, and a strong rivalry emerged. Up to 1617, Willem Blaeu signed his work Guilielmus Janssonius or Willem Jans Zoon, while his later work was signed G. Blaeu / Willem Blaeu. In fact, in 1621 Willem Janszoon added the surname Blaeu(w) (sometimes in the latinized form Caesius) to his imprints. That had almost certainly to do with the rivalry between him and Johannes Janssonius, whose name strongly resembled. By the addition of the name Blaeu, confusion was avoided.

A strange episode took place in 1630, with the sale of 37 copperplates of the *Mercator Atlas* from Jodocus Hondius Jr. by his widow to Willem Jansz. Blaeu, the most important competitor of the Hondius-Janssonius firm ! Blaeu replaced Jodocus Hondius Jr's name with his own on the plates and the following year, he had published them together with his own maps in the *Atlantis Appendix*, which contained 60 maps. Five years later he issued the first two volumes of his planned world atlas, *Atlas Novus* or the *Theatrum Orbis Terrarum*.

In 1633, the States General of Amsterdam appointed Blaeu map maker of the Republic, and later he became the official cartographer of the Dutch East India Company. About the same time he was appointed Hydrographer to the Dutch East India Company, Vereenighde Oostindische Compagnie (VOC). During this period, Amsterdam was one of the wealthiest trading cities in Europe and a center for banking and the diamond trade. The VOC contributed significantly to the wealth and prosperity of the United Netherlands, and Blaeu's prestigious appointment firmly established his reputation within the highly competitive field of Dutch mapmakers.

Willem Blaeu's intention to publish a new "international edition" of a world atlas is mentioned on 11 February 1634 in an Amsterdam newspaper, the *Courante uyt Italien ende Duytschlandt*, published by Jan van Hilten. The atlas, entitled *Theatrum Orbis Terrarum*, was first published in 1634 in German, then in 1635 in Latin, Dutch and French.

Blaeu's plans were ambitious. In the preface to the 1636 German edition, he wrote : "*Our intention is to describe the entire world, that is the heavens and the earth, in several volumes like these two, of which two more of the earth will shortly follow*".

But Willem Jansz. Blaeu did not live to see the other two volumes issued which he had prepared. When he died on the 21st of October 1638, the business passed to his sons Joan (1596-1673) and Cornelis (c. 1610-1644), who continued and expanded their father's ambitious plans. The two additional volumes appeared in 1640 in Italy and in 1645 in England.

After the death of Cornelis in 1644, Joan established his own reputation as a great mapmaker. He completed his father grand project in 1655 with the sixth volume of the atlas, and in 1662 the most voluminous and magnificent world atlas of all times, the *Atlas Major* (in 9 or 12 volumes according to the Dutch or French editions) saw the light.

The *Atlas Major* was the most expensive printed book of the seventeenth century, consisting of nearly 600 double-page maps and 3,000 pages of text. The maps were richly embellished, often hand-colored and heightened with gold, and epitomized the style and quality of the period, which has become known as the Golden Age of cartography. In addition to geographical maps, the *Atlas Major* describes and illustrates Brahe's astronomical instruments. Blaeu's handcolored copper-plate engravings were revised from woodcuts originally published in Brahe's own *Astronomiae Instauratiae Mechanicae* (1598), with the descriptions in Latin ; the section also gives a map of Hven island and plans and descriptions of Tycho's two observatories, Uraniborg and Stelleborg.

The last eleventh volume of the Spanish edition was in the press when, on 28 February 1672, the printing shop was destroyed by fire, a blow from which the firm of Blaeu never recovered.

**Reference books**

*Nieuw nederlandsch biografisch woordenboek*, **10**, 74-8 (1937), Leiden : A.W. Sijthoff.

E. L. Stevenson, *Willem Janszoon Blaeu, a sketch of his life and work*, New York : De Vinne Press, 1914.

G. Schilder, *Monumenta Cartographica Neerlandica*, vol. 3, Alphen a/d Rijn, 1987.

*Koeman's Atlantes Neerlandici*, Revised edition by Peter van der Krogt, vol II : *The Folio Atlases Published by Willem Jansz. Blaeu and Joan Blaeu*. Utrecht , 2000 - ISBN 9789061944386

J. Keuning, *Willem Jansz. Blaeu : A biography and history of his work as a cartographer and publisher*, revised and edited by Marijke Donkersloot-De Vrij. Amsterdam : Theatrum Orbis Terrarum, 1973.  ISBN 9022112535

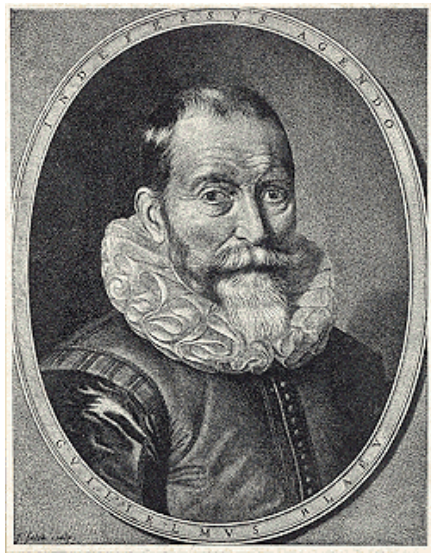

Portrait of W. J. Blaeu by Jeremias Falck, engraving from the *Digitale bibliootheek voor de Nederlandse letteren* (Public domain)